\documentclass[12pt]{iopart}
\usepackage{booktabs}
\usepackage{graphicx}
\usepackage{color}

\newcommand{\KFeAs}{KFe$_2$As$_2$}

\newcommand{\p}[1]{\left( #1 \right)}

\begin{document}

\title{From $d$-wave to $s$-wave pairing in the iron-pnictide superconductor (Ba,K)Fe$_2$As$_2$}

\author{J.-Ph. Reid$^1$, A.~Juneau-Fecteau$^1$, R.~T.~Gordon$^1$, S.~Ren\'e~de~Cotret$^1$, N.~Doiron-Leyraud$^1$, 
X.~G.~Luo$^1$, H.~Shakeripour$^1$, J.~ Chang$^1$,
M.~A.~Tanatar$^{2,3}$, H.~Kim$^{2,3}$, R.~Prozorov$^{2,3}$,
T.~Saito$^4$, H.~Fukazawa$^4$, Y.~Kohori$^4$,
K.~Kihou$^5$, C.~H.~Lee$^5$, A.~Iyo$^5$, H.~Eisaki$^5$,
B. Shen$^6$, H.-H. Wen$^{6,7}$,
and Louis Taillefer$^{1,7}$}
\address{
$^1$D\'epartement de physique \& RQMP, Universit\'e de Sherbrooke, Sherbrooke, Qu\'ebec, Canada J1K 2R1\\
}
\address{
$^2$Ames Laboratory, Ames, Iowa 50011, USA\\
}
\address{
$^3$Department of Physics \& Astronomy, Iowa State University, Ames, Iowa 50011, USA\\
}
\address{
$^4$Chiba University \& JST-TRIP, Japan\\
}
\address{
$^5$AIST \& JST-TRIP, Japan\\
}
\address{
$^6$Center for Superconducting Physics and Materials, National Laboratory of Solid State Microstructures
and Department of Physics, Nanjing University, Nanjing 210093, China\\
}
\address{
$^7$Canadian Institute for Advanced Research, Toronto, Ontario, Canada M5G 1Z8
}
\ead{louis.taillefer@usherbrooke.ca}

\begin{abstract}
The nature of the pairing state in iron-based superconductors is the subject of much debate.
Here we argue that in one material, the stoichiometric iron pnictide KFe$_2$As$_2$,
there is overwhelming evidence for a $d$-wave pairing state, characterized by
symmetry-imposed vertical line nodes in the superconducting gap.
This evidence is reviewed, with a focus on thermal conductivity and the 
strong impact of impurity scattering on the critical temperature $T_c$.
We then compare KFe$_2$As$_2$ to Ba$_{0.6}$K$_{0.4}$Fe$_2$As$_2$, obtained
by Ba substitution, where the pairing symmetry is $s$-wave and the $T_c$ is
ten times higher. 
The transition from $d$-wave to $s$-wave within the same crystal structure provides 
a rare opportunity to investigate the connection between band structure and pairing mechanism.
We also compare KFe$_2$As$_2$ to the nodal iron-based superconductor LaFePO,
for which the pairing symmetry is probably not $d$-wave,
but more likely $s$-wave with accidental line nodes.

\end{abstract}
\maketitle
\section{Introduction}  

Iron pnictides show that high-temperature superconductivity can be achieved in a good metal on the border of antiferromagnetic order.
A key question is the role of the antiferromagnetic spin fluctuations in causing the pairing.
Early on it was proposed that the two-band nature of the materials (Fig.~1), with one (or more) hole-like Fermi surface at the zone center
($\Gamma$ point) and an electron-like Fermi surface at the zone corner ($X$ point), is responsible for both the spin-stripe order and the superconductivity \cite{Mazin2008}.
Pairing would rely predominantly on the inter-band interaction that links the hole and electron Fermi surfaces, producing a pairing state 
with $s$-wave symmetry but with a gap that has opposite sign on the two Fermi surfaces, the so-called $s_{\pm}$ state.

In this Article we consider a particular iron pnictide, KFe$_2$As$_2$, which offers an unprecedented 
window on the workings of magnetically-mediated superconductivity, as it undergoes a phase transition from one 
pairing symmetry to another, induced simply by substitution of Ba on the K site. 
The change from $d$-wave symmetry in KFe$_2$As$_2$ 
to $s$-wave symmetry in Ba$_{1-x}$K$_x$Fe$_2$As$_2$ is accompanied by 
the appearance of the electron pocket in the Fermi surface
and a ten-fold increase in $T_c$ (see Fig.~2).
This change of symmetry was predicted theoretically by renormalization-group calculations~\cite{Thomale2011} and 
recently confirmed by experiment~\cite{Reid2012}. 
Here we review the evidence.


\begin{figure}[t]
\centering
\includegraphics[width=12cm]{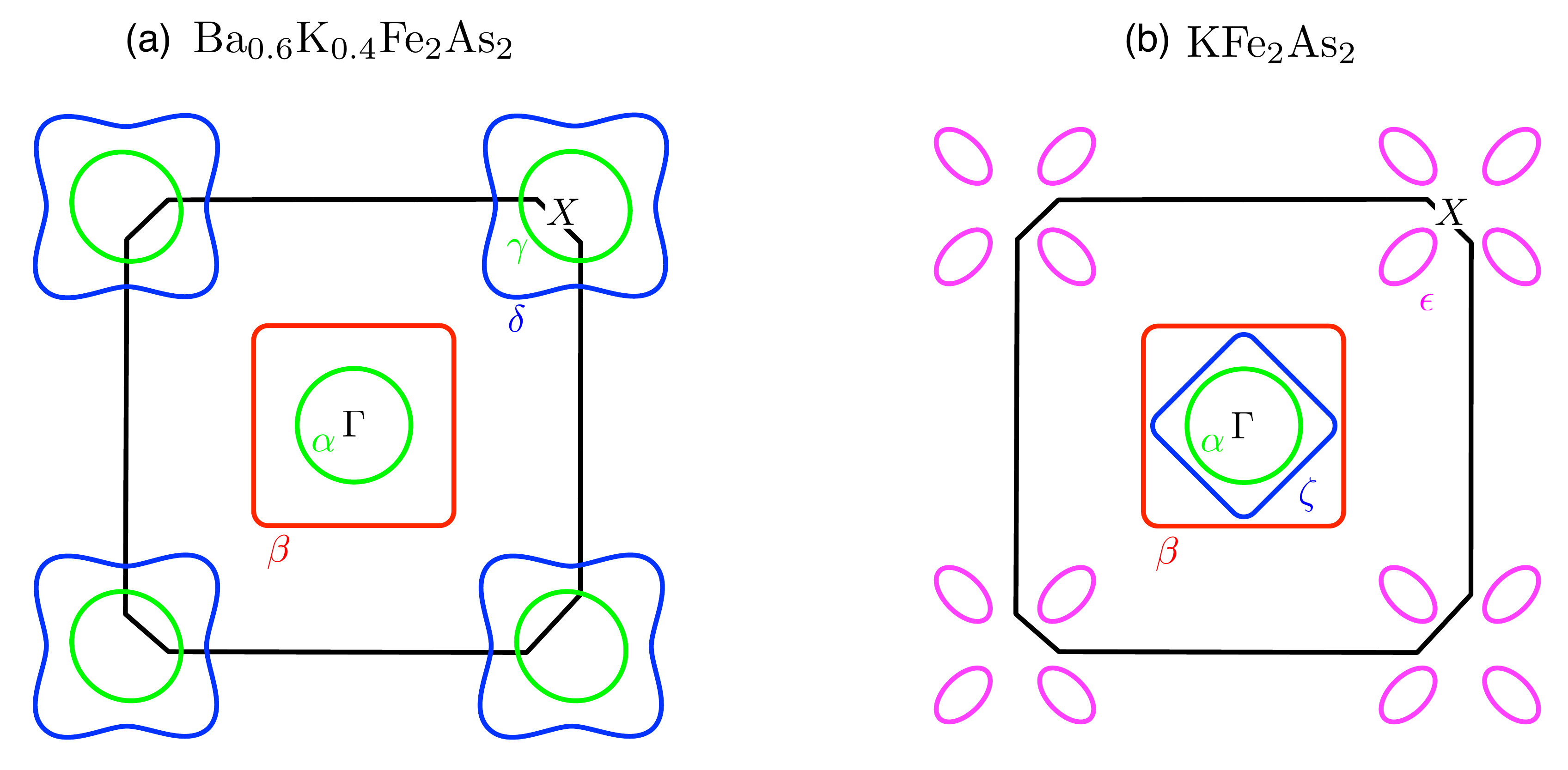}
\caption{
Sketch of the Fermi surface of the iron pnictide Ba$_{1-x}$K$_x$Fe$_2$As$_2$ in the $k_z=0$ plane, for two K concentrations.
(a) For $x=0.4$:
two hole-like cylinders at the zone center ($\Gamma$) and two electron-like cylinders at the zone corner ($X$)\cite{Xu2008}.
(b) For $x=1.0$:
three hole-like cylinders at the zone center ($\Gamma$), one hole-like cylinder near the zone corner ($X$), and
no electron pocket~\cite{Hashimoto2010a,Sato2009}. 
Recent ARPES data~\cite{Yoshida2012} confirms this Fermi surface topology in the $k_z = 0$ plane; it also reveals the presence of a small
3D pocket centered at the $Z$ point of the Brillouin zone (at $k_z = \pi/c$, not shown), consistent with calculations~\cite{Hashimoto2010a}.
The disappearance of the electron pocket in going from $x=0.4$ to $x=1.0$ is thought to play a major role in the dramatic change
in the superconducting state: from $s$-wave symmetry with $T_c \simeq 40$~K to $d$-wave symmetry with $T_c \simeq 4$~K.
}
\label{Fig6}
\end{figure}



\begin{figure}[t]
\centering
\includegraphics[width=8.5cm]{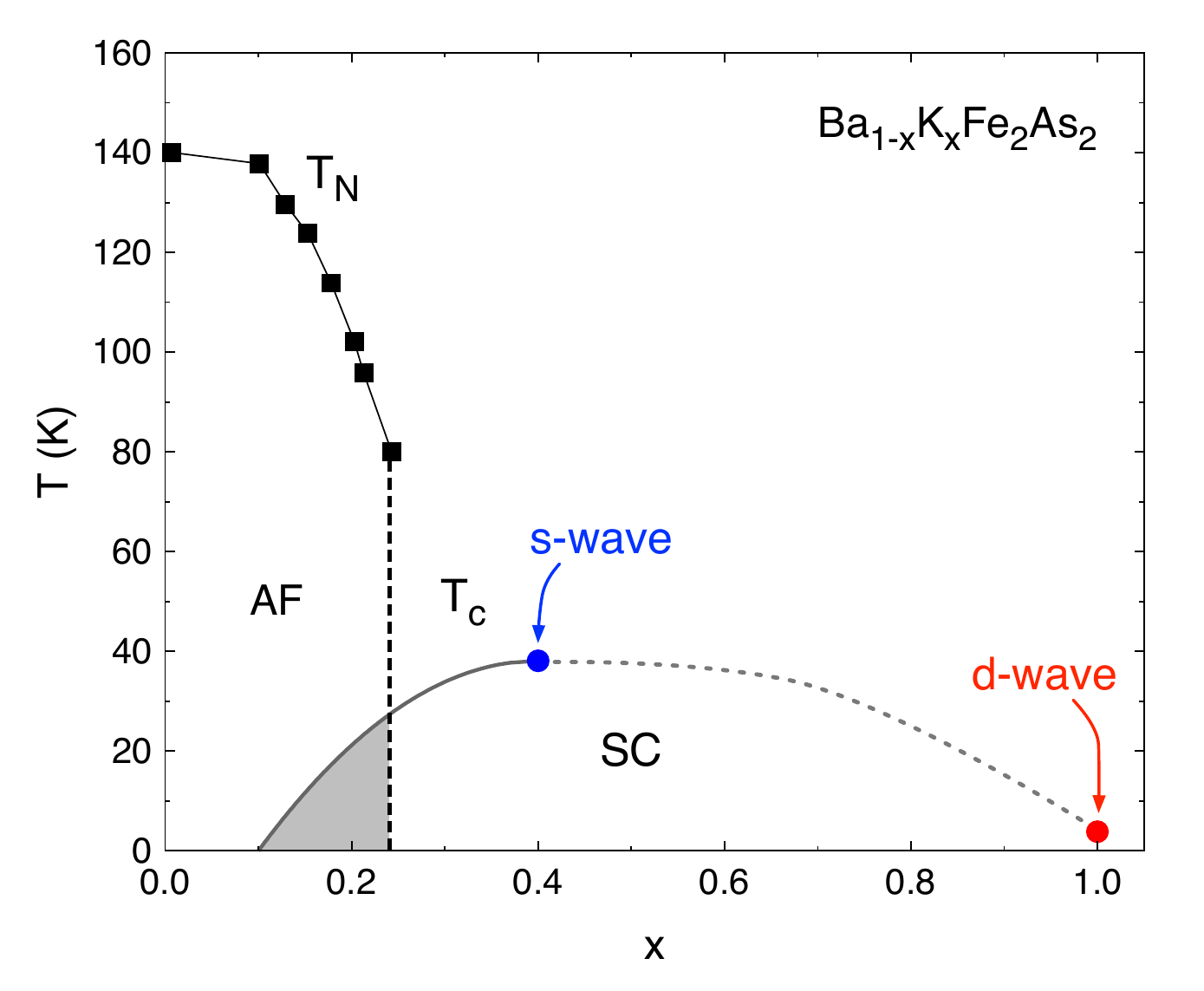}
\caption{
Phase diagram of the iron-pnictide superconductor Ba$_{1-x}$K$_x$Fe$_2$As$_2$ as a function of K concentration $x$.
At $x=0$, the parent compound is antiferromagnetic (AF) below a temperature $T_{\rm N}$ which
decreases with $x$ (black squares~\cite{Avci2011}). 
Superconductivity exists in a dome-like region below the transition temperature $T_c$.
With increasing $x$, $T_c$ rises to reach a maximal value of 38~K at optimal doping ($x \simeq 0.4$), outside the region
of coexisting antiferromagnetic order (shaded in grey).
%
%
At $x=1.0$, single crystals have the lowest disorder of any iron-based superconductor.
The properties of KFe$_2$As$_2$ are very well characterized, and $T_c = $~4.0 K in the cleanest samples. 
In this article, we compare the system at two different concentrations:
$x = 0.4$ (blue dot)
and $x=1.0$ (red dot).
We show that the symmetry of the superconducting state changes from $s$-wave at $x = 0.4$ to $d$-wave at $x=1.0$. 
This means that there must be a phase transition in between.
%
%
%
%
%
%
%
}
\label{PhaseDiagramComplet}
\end{figure}


\subsection{$d$-wave vs $s$-wave}  

There are two kinds of nodes where a superconducting gap goes to zero: 
those that are imposed by the symmetry of the pairing state, and those that are accidental,
not imposed by any symmetry but the result of an anisotropic pairing interaction. 
In a $d$-wave state, of $B_{1g}$ symmetry in a tetragonal structure, the gap goes to zero along diagonals in the $x$-$y$ plane.
In the simplest 2D model, with $\Delta = \Delta_0$cos$2\phi$, $\Delta(\phi)=0$ when the azimuthal angle $\phi$ equals $\pi/4$,
$3\pi/4$, $5\pi/4$ and $7\pi/4$. The order parameter changes sign as $\phi$ goes from 0 to $\pi/2$ and so it breaks the
tetragonal symmetry of the lattice. 
If the Fermi surface is a quasi-2D cylinder located at the center of the Brillouin zone, then the gap 
will necessarily have 4 vertical line nodes running parallel to the axis of the cylinder. See Fig.~3.

An $s$-wave gap has $A_{1g}$ symmetry and so preserves the rotational symmetry of the underlying lattice.
It can have accidental nodes, but the number of nodes where the gap changes sign must preserve tetragonal symmetry. 
On a quasi-2D Fermi surface cylinder in a tetragonal structure, an $s$-wave gap with nodes (called an "extended $s$-wave" gap)
can go to zero along lines that can either be vertical or horizontal. 
For example, the gap can have 4 pairs of vertical line nodes, with 4 negative minima separating 4 positive maxima, as
sketched in Fig.~3.


\begin{figure}[t]
\centering
\includegraphics[width=8.5cm]{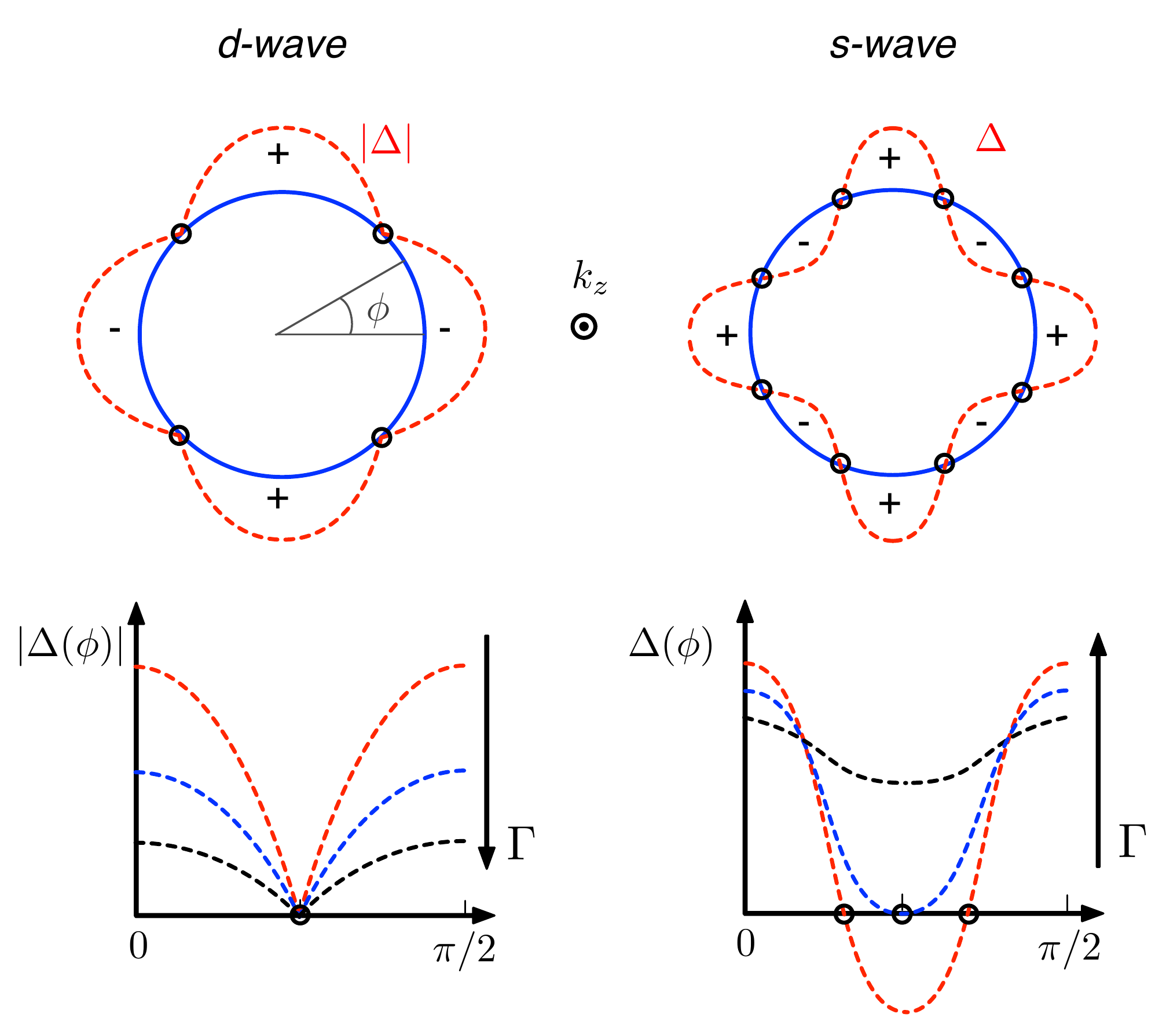}
\caption{
{\it Top panels}:
Sketch of the superconducting gap $\Delta(\phi)$ (red dashed line) as a function of azimuthal angle $\phi$
in the basal plane of a tetragonal quasi-2D superconductor with a circular Fermi surface (blue solid line).
A $d$-wave gap (on the left) changes sign 4 times as it goes around the Fermi surface 
and as a result the gap necessarily goes to zero at four nodes (small black circles).
It breaks the four-fold rotational symmetry of the lattice.
An $s$-wave gap (right) does not break the four-fold rotational symmetry of the lattice.
In general it does not change sign, but it can do so, at accidental nodes, as sketched here for example.
{\it Bottom panels}:
Effect of increasing the impurity scattering rate $\Gamma$. 
For a $d$-wave gap (left), the nodes always remain, as they are imposed by the symmetry of the order parameter.
Impurity scattering rapidly suppresses the gap magnitude.
For an $s$-wave gap (right), scattering makes the gap more isotropic and will in general remove the nodes.
It is less effective in reducing the gap magnitude.
}
\label{Fig6}
\end{figure}


\section{Effect of impurity scattering on $T_c$}  

Impurity scattering has a qualitatively different effect on $d$-wave and extended $s$-wave states, 
as pointed out early on in the study of cuprate superconductors~\cite{Borkowski1994}.
Scattering cannot remove the $d$-wave nodes, as they are imposed by symmetry.
By contrast, scattering in the $s$-wave state will in general make the gap more isotropic, 
and eventually remove (lift) the nodes (Fig.~3).

Intra-band non-magnetic scattering is very effective in suppressing the $T_c$ of a $d$-wave superconductor.
The scattering rate $\Gamma$ needed to suppress $T_c$ to zero is of the order of $k_{\rm B} T_{c0} / \hbar$, where $T_{c0}$ is the clean-limit value of $T_c$.
In a single-band model with unitary scattering, the critical scattering rate $\Gamma_c$ is given by 
$\hbar \Gamma_c = 0.88~k_{\rm B} T_{c0}$~\cite{SunMakiEPL1995,Alloul2009}.
Intra-band scattering is much less effective in suppressing the $T_c$ of an $s$-wave superconductor~\cite{Mishra2009}.

In this context, let us examine the behaviour of iron-based superconductors.
In Fig.~4, we plot $T_c$ vs $\Gamma$, both normalized by $T_{c0}$, for various 122 superconductors.
In KFe$_2$As$_2$, $\hbar \Gamma_c  \simeq k_{\rm B} T_{c0}$, in agreement with 
expectation for a $d$-wave state.
$\Gamma_c$ is estimated using the critical value of the residual resistivity $\rho_0$ for which 
$T_c$ will go to zero, namely $\rho_0^{\rm crit} \simeq 4.5~\mu \Omega$~cm in KFe$_2$As$_2$~\cite{Reid2012}.
%
%
Using the normal-state specific heat coefficient 
$\gamma_{\rm N} = 85 \pm 10$~mJ/K$^2$ mol~\cite{Abdel-Hafiez2011,Fukazawa2011} and the average Fermi velocity
$v_{\rm F} \simeq 4 \times 10^4$~m/s~\cite{Terashima2010a}, we obtain
$\hbar \Gamma_c = \hbar /2\tau_c  \simeq 1.3 \pm 0.2~k_{\rm B} T_{c0}$,
via
the normal-state thermal conductivity
$\kappa_{\rm N}/T = L_0 / \rho_0^{\rm crit} = \gamma_{\rm N} v_{\rm F}^2 \tau_c / 3$,
where $L_0 \equiv (\pi^2 / 3) (e / k_{\rm B})^2$.


\begin{figure}[t]
\centering
\includegraphics[width=8.5cm]{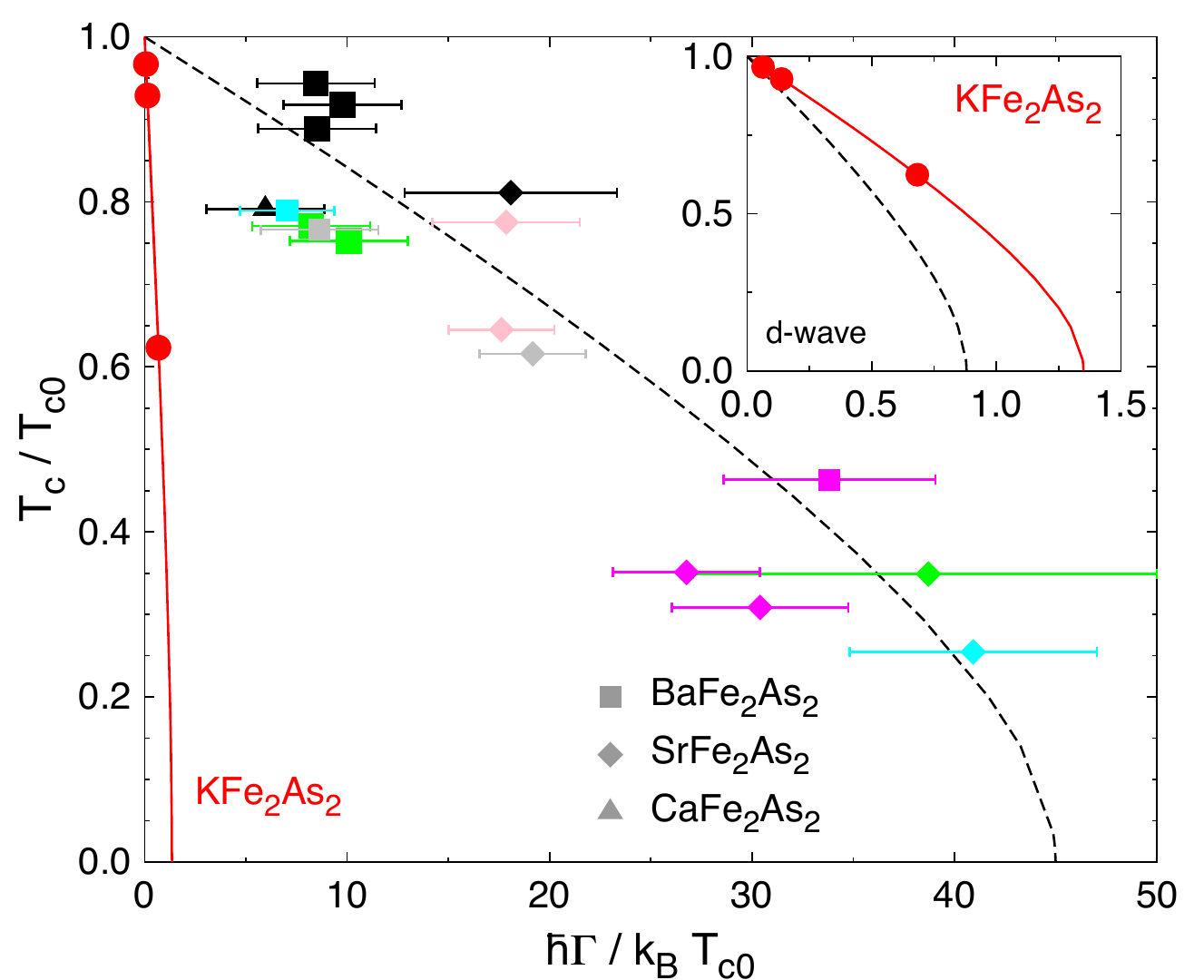}
\caption{
Effect of impurity scattering on the transition temperature $T_c$ of various iron-pnictide superconductors. 
The scattering rate $\Gamma$ is estimated using the residual resistivity $\rho_0$.
%
%
Data for BaFe$_2$As$_2$ (squares), 
SrFe$_2$As$_2$ (diamonds)
and CaFe$_2$As$_2$ (triangles) optimally doped with Co, Pt or Ru are reproduced from 
ref.~\cite{Kirshenbaum2012}. 
Data for KFe$_2$As$_2$ (red circles) are reproduced from ref.~\cite{Reid2012}.
%
%
%
Both axes are normalized by $T_{c0}$, the value of $T_{c}$ in the clean limit ($\Gamma \to 0$);
$T_{c0} = 3.95$~K for KFe$_2$As$_2$~\cite{Reid2012} 
and 
$T_{c0} = 26$~K for optimally doped BaFe$_2$As$_2$, SrFe$_2$As$_2$ and CaFe$_2$As$_2$~\cite{Kirshenbaum2012}.
All lines are the same theoretical Abrikosov-Gorkov curve for the decrease of $T_c$ with scattering rate,
plotted as $T_c / T_{c0}$ vs $\Gamma / \Gamma_c$, where the critical scattering rate $\Gamma_c$ 
for suppressing $T_c$ to zero is the only adjustable parameter. 
For KFe$_2$As$_2$, $\hbar \Gamma_c = 1.3~k_{\rm B} T_{c0}$ (see text), while for the other compounds,
$\hbar \Gamma_c = 45~k_{\rm B} T_{c0}$.
The 30-fold difference in $\Gamma_c / T_{c0}$ is compelling evidence for superconducting states of different symmetry. 
{\it Inset :}
Zoom on the KFe$_2$As$_2$ data at small $\Gamma$. 
The black dashed line is the theoretical expectation for a $d$-wave superconductor,
assuming a single-band Fermi surface, for which
$\hbar \Gamma_c = 0.88~k_{\rm B} T_{c0}$~\cite{SunMakiEPL1995,Alloul2009}.
}
\label{Fig6}
\end{figure}



Note that in this and other estimates below, we use single-band models and relations to compare theory and experiment in KFe$_2$As$_2$,
a multi-band material. 
The reason why a single-band approximation may work reasonably well, is that the Fermi surface sheets that dominate the spectral weight
are the three relatively similar hole-like cylinders centered on the $\Gamma$ point (Fig.~1)~\cite{Hashimoto2010a,Terashima2010a}.
The weak magneto-resistance of KFe$_2$As$_2$ was offered as evidence of single-band behavior~\cite{Terashima2009}.


In dramatic contrast, $T_c$ is suppressed 30 times more slowly
in the materials based on BaFe$_2$As$_2$, SrFe$_2$As$_2$ and CaFe$_2$As$_2$, optimally doped
by substitution on the Fe site (with Co, Co/Cu, Pt, Ru, Ni, and Pd).
Indeed, all of those were found to fall on the same curve,
with $T_c$ going to zero at $\hbar \Gamma_c \simeq 45~k_{\rm B} T_{c0}$~\cite{Kirshenbaum2012}. 
This is completely incompatible with $d$-wave pairing, and indeed some of these materials have gaps without nodes,
{\it e.g.} Co-doped BaFe$_2$As$_2$~\cite{Tanatar2010,Reid2010}.
This 30-fold difference between KFe$_2$As$_2$ and materials that otherwise have an identical crystal structure and a very similar 
Fermi surface is compelling evidence that the pairing symmetry must be different.
%
We now use heat transport at very low temperature to show that the pairing symmetry of KFe$_2$As$_2$ is indeed $d$-wave.


\section{Theory of heat transport in a $d$-wave superconductor}  

The thermal conductivity $\kappa$ measured at very low temperature is a powerful probe of the superconducting state~\cite{Shakeripour2009}.
It is a bulk measurement, insensitive to modifications that might occur at the surface of a sample.
It is directional, and so can access the anisotropy of the gap structure.
And it probably has the best energy resolution of all experimental probes for distinguishing nodes in the gap from deep minima, dictated by the lowest temperature
of the measurement, typically 50 mK or so.
Finally, as we shall see, it can distinguish between symmetry-imposed line nodes and accidental line nodes.

The presence of a residual linear term in $\kappa(T)$, whereby $\kappa/T$ is non-zero in the limit of $T \to 0$,
is unambiguous evidence of zero-energy fermionic quasiparticles. 
In a metal or in the normal state of a superconductor induced for example by applying a magnetic field greater than $H_{c2}$, 
this residual linear term, labelled $\kappa_{\rm N}/T$, is given by the Wiedemann-Franz law :
\begin{equation}
\frac{\kappa_{\rm N}}{T} = \frac{L_0}{\rho_0}  ~~~~~~.
\end{equation}
%

In the superconducting state at zero magnetic field ($H=0$), a non-zero residual linear term, labelled $\kappa_0/T$,
is unambiguous evidence that the superconducting gap must vanish somewhere on the Fermi surface.
The gap must have nodes, whether points or lines, accidental or symmetry-imposed.
By contrast, if the gap does not have nodes, even if it might have deep minima, then 
$\kappa_0/T = 0$, as confirmed by measurements in the $s$-wave superconductors V$_3$Si~\cite{Sutherland2003}, MgB$_2$~\cite{Sologubenko2002}, 
NbSe$_2$~\cite{Boaknin2003} and LuNi$_2$B$_2$C~\cite{Boaknin2001},
for example.

The theory of heat transport in a $d$-wave superconductor is well 
developed~\cite{SunMakiEPL1995,Graf1996,Graf1996a,Kubert1998,Vekhter1999,Durst2000}. 
The zero-energy quasiparticles
associated with the nodes in the $d$-wave gap produce fermionic heat transport in the $T \to 0$ limit.
These quasiparticles are induced by impurity scattering, and further excited by a magnetic field
or a rise in temperature, as we shall now describe.
%

\subsection{Magnitude of $\kappa_0/T$ and its dependence on impurity scattering}

Calculations show that the residual linear term in a $d$-wave superconductor is independent of scattering rate and phase shift~\cite{Graf1996}, 
and free of Fermi-liquid and vertex corrections~\cite{Durst2000}.
For a quasi-2D superconductor in the clean limit ($\hbar \Gamma \ll \Delta_{0}$), it is given by~\cite{Graf1996,Durst2000}:
\begin{equation}
\frac{\kappa_{00}}{T}~ \simeq ~ \frac{k_{\rm B}^2}{3\hbar} ~\frac{1}{c} ~ \frac{v_{\rm F}}{v_\Delta}~~~~~~,
\end{equation}
where $c$ is the interlayer separation,
$v_{\rm F}$ is the
in-plane Fermi velocity,
and $v_\Delta$ is the quasiparticle velocity tangential to the Fermi surface at the node,
proportional to the slope of the gap at the node.
We will use this expression to estimate the theoretically expected value of $\kappa_0/T$ for the cuprate superconductor
YBa$_2$Cu$_3$O$_7$.

Eq.~2 may also be expressed in terms of the residual linear term in the normal-state electronic specific heat, $\gamma_{\rm N}$,
and the slope of  the gap at the node, $\mu \Delta_0$ (expressed in terms of the gap maximum, $\Delta_0$)~\cite{Graf1996}:
\begin{equation}
\frac{\kappa_{00}}{T} ~\simeq ~ \frac{\hbar}{\pi}~  \frac{\gamma_{\rm N} v^2_{\rm F}}{\mu \Delta_0}~~~~~~.
\end{equation}
In the simple case where $\Delta = \Delta_0$cos($2 \phi$), $\mu = 2$.  
Assuming the weak-coupling value of the gap, $\Delta_0 = 2.14~k_{\rm B} T_{c0}$, and $\mu =2$, we get:
\begin{equation}
\frac{\kappa_{00}}{T} ~= ~\frac{\hbar}{4.28 \pi}  ~\frac{\gamma_{\rm N} v^2_{\rm F}}{k_{\rm B} T_{c0}}~~~~~~.
\end{equation}
We will use this expression to estimate the theoretically expected value of $\kappa_0/T$ for KFe$_2$As$_2$.

The remarkable feature of $d$-wave superconductors (or indeed any superconductor with
symmetry-imposed line nodes) is that $\kappa_0/T$ is  independent of the impurity scattering rate $\Gamma$.
This comes from a compensation between the impurity-induced growth in quasiparticle density and the
proportional decrease in mean free path. 
This is what is meant by ``universal heat conduction", first verified experimentally in the cuprate superconductor YBa$_2$Cu$_3$O$_7$~\cite{Taillefer1997}. 
Although it is strictly true only in the clean limit ($\hbar \Gamma \ll \Delta_{0}$),
the change in $\kappa_0/T$ is expected to remain weak up to $\hbar \Gamma / k_{\rm B} T_{c0} \simeq 0.5$~\cite{SunMakiEPL1995}.

\subsection{Dependence of $\kappa/T$ on temperature}  

So far, we have discussed the limit $T \to 0$ and $H \to 0$, where nodal quasiparticles are excited only by the pair-breaking effect of impurities.
Raising the temperature will further excite nodal quasiparticles.
Calculations for a $d$-wave superconductor show that the leading-order finite-temperature correction to Eq.~2 (or Eq.~3)
is a positive contribution to the electronic thermal conductivity that
grows as $T^2$~\cite{Graf1996,Graf1996a}:
\begin{equation}
\frac{\kappa}{T} ~ \simeq ~  \frac{\kappa_{00}}{T} ~~( 1 + a ~ \frac{T^2}{\gamma^2} )~~~~~~~~~,
\end{equation}
where $a$ is a dimensionless number and $\gamma$ is the impurity bandwidth,
which grows with the scattering rate $\Gamma$~\cite{Graf1996}.
In the limit of unitary scattering, $\gamma^2 \propto \Gamma$; away from unitary scattering, towards
Born scattering, $\gamma^2$ grows more rapidly with $\Gamma$~\cite{Graf1996}.

\subsection{Dependence of $\kappa_0/T$ on magnetic field}

Increasing the magnetic field is another way to excite quasiparticles in the superconducting state.
In the absence of nodes, the quasiparticle states at $T=0$ are localized in the vortex core and heat transport occurs only as a result 
of tunneling between adjacent vortices~\cite{Shakeripour2009}.
The dependence of $\kappa_0/T$ on $H$ is then exponentially slow. 
If the gap has nodes, however, the field will shift the energy of delocalized states outside the vortex cores
and enhance the residual density of states, which will grow typically as $\sqrt{H}$.
This causes an immediate rise in $\kappa_0/T$ with applied field~\cite{Shakeripour2009,Kubert1998,Vekhter1999}.
Calculations for a quasi-2D $d$-wave superconductor in the clean limit ($\hbar \Gamma \ll \Delta_0$)
reveal a non-monotonic increase of $\kappa_0/T$ vs $H$, with most of the increase to reach 
$\kappa_{\rm N}/T$ occurring in the last quarter of $H/H_{c2}$~\cite{Vekhter1999} (see Fig.~9).
%


\section{Experimental data for KFe$_2$As$_2$}  

In this section, we compare the properties of KFe$_2$As$_2$ to those expected of a $d$-wave superconductor,
both from theory, as outlined in the previous section, and from observation in cuprates, the archetypal $d$-wave superconductors.
We also compare to theoretical calculations for an extended $s$-wave gap applied to pnictides~\cite{Mishra2009}.
Finally, we compare with the equivalent data on Ba$_{0.6}$K$_{0.4}$Fe$_2$As$_2$.

We begin by showing, in Fig.~5, thermal conductivity data for KFe$_2$As$_2$ and Ba$_{0.6}$K$_{0.4}$Fe$_2$As$_2$,
at $H \simeq 0$.
(A very small field $H = 0.05$~T is applied to kill the superconductivity of the contacts~\cite{Reid2010}.)
It is clear by inspection that there is a large residual linear term in KFe$_2$As$_2$ and none in Ba$_{0.6}$K$_{0.4}$Fe$_2$As$_2$.
The fact that  $\kappa_0/T = 0$ in the latter~\cite{Reid2011} is sufficient proof to rule out $d$-wave symmetry for the superconducting state at $x=0.4$.


\begin{figure}[t]
\centering
\includegraphics[width=8.5cm]{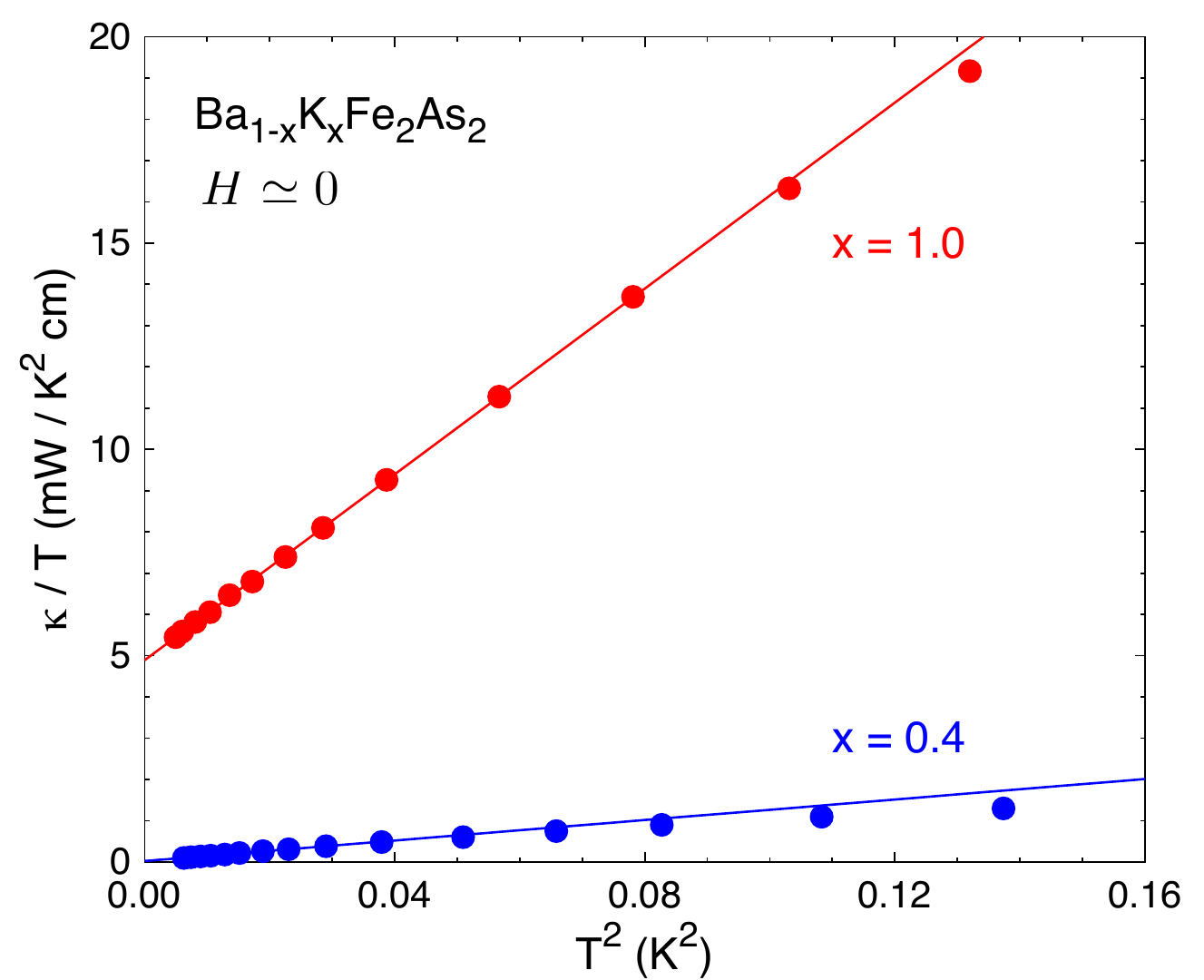}
\caption{
Temperature dependence of the in-plane thermal conductivity $\kappa$ of Ba$_{1-x}$K$_x$Fe$_2$As$_2$,
measured in a very small magnetic field ($H \simeq 0$), 
plotted as $\kappa/T$ vs $T^2$, for two K concentrations:
$x=0.4$ (blue) \cite{Reid2011} and $x=1.0$ (red)~\cite{Reid2012}. 
The lines are a linear fit, used to extrapolate the residual linear term at $T=0$, labelled $\kappa_0/T$. 
}
\label{Fig2}
\end{figure}


\subsection{Magnitude of $\kappa_0/T$}  

When extrapolated to zero field (see inset of Fig.~9), the residual linear term in KFe$_2$As$_2$ is~\cite{Reid2012}:
\begin{equation}
\frac{\kappa_{0}}{T} ~ =  ~3.6 \pm 0.5~{\rm mW/K}^2~{\rm cm}~~~.
\end{equation}
Let us compare this with the value expected for a $d$-wave gap.
Using Eq.~4 and $\gamma_{\rm N} v^2_{\rm F} = 2.3 \pm 0.3 \times 10^{13}$~mJ/K$^2$ cm s$^2$~\cite{Reid2012},
we get
\begin{equation}
\frac{\kappa_{00}}{T} ~ =  ~3.3 \pm 0.5~{\rm mW/K}^2~{\rm cm}~~~,
\end{equation}
in excellent agreement with experiment.

Let us also compare with experimental data on cuprate superconductors.
Measurements on YBa$_2$Cu$_3$O$_7$ near optimal doping, at a doping $p\simeq0.18$ where $T_c = 89$~K give~\cite{Hill2004}:
\begin{equation}
\frac{\kappa_{0}}{T} ~ =  ~0.16 \pm 0.02~{\rm mW/K}^2~{\rm cm}~~~.
\end{equation}
Using Eq.~2 and the ratio of quasiparticle velocities measured by ARPES in Ba$_2$Sr$_2$CaCu$_2$O$_{8+\delta}$, 
$v_{\rm F}/v_\Delta = 16 \pm 1$ at $p \simeq 0.16$~\cite{Vishik2010}, yields
\begin{equation}
\frac{\kappa_{00}}{T} ~ =  ~ 0.16 \pm 0.01~{\rm mW/K}^2~{\rm cm}~~~,
\end{equation}
since the interlayer separation $c = 5.85$~\AA~in YBa$_2$Cu$_3$O$_7$.
This is in perfect agreement with experiment.
The same is true for Tl$_2$Ba$_2$CuO$_{6+\delta}$  at $p=0.19$ with $T_c = 84$~K: the measured value,
$\kappa_{0}/{T} = 0.08 \pm 0.02~{\rm mW K}^2~{\rm cm}$~\cite{Hawthorn2007}, is half that in YBCO 
because the interlayer separation, $c=11.6$~\AA, is twice what it is in YBCO.

It is important to appreciate, however, that Eqs. 2 and 3 are only valid in the clean limit, when
$\hbar \Gamma \ll \Delta_{0}$. 
This condition holds well for the cleanest samples of KFe$_2$As$_2$, with RRR values of 1000 or so~\cite{Reid2012}.
An easy way to quickly estimate the ratio $\hbar \Gamma / \Delta_{0}$ is via the ratio of
superconducting-state to normal-state residual linear terms:
$(\kappa_{0}/T) / (\kappa_{\rm N}/T) \simeq \hbar \Gamma / \Delta_0$.
For a sample of KF$_2$As$_2$ with $\rho_0 = 0.2~\mu \Omega$~cm, this gives
$\hbar \Gamma / \Delta_0 = 0.025$. 
The clean-limit condition almost certainly also holds for the cuprates YBa$_2$Cu$_3$O$_y$ and Tl$_2$Ba$_2$CuO$_{6+\delta}$ near optimal doping,
but it doesn't in strongly overdoped Tl$_2$Ba$_2$CuO$_{6+\delta}$~\cite{Proust2002}.
Neither does it hold for most other cuprates, including Ba$_2$Sr$_2$CaCu$_2$O$_{8+\delta}$~\cite{Sun2006}
and La$_{1-x}$Sr$_x$CuO$_4$~\cite{Sutherland2003,Sun2006} (see discussion in~\cite{Li2008}).
In the dirty limit, the thermal conductivity is no longer a simple measure of the $d$-wave gap near the node,
but instead it reflects the normal-state character of the conductivity.

These quantitative comparisons with theory show how reliable the absolute values of
$\kappa_0/T$ can be in assessing whether a particular superconductor has a $d$-wave gap. 
There are two reasons for this. First, $\kappa_0/T$ does not depend on the
concentration or type of impurities or defects in the sample. Secondly, there is typically a direct
connection between the slope of the gap at the node, $v_\Delta$, which controls the density of
excited zero-energy quasiparticles, and the gap maximum $\Delta_0$, and hence $T_c$.
Neither of these reasons applies to the case of accidental line nodes in an extended $s$-wave
gap. Indeed, while Eq.~2 still holds~\cite{Mishra2009}, $v_\Delta$ is now strongly dependent on the impurity scattering 
and its value is not related in any simple way to $T_c$. 
So the remarkable agreement in KFe$_2$As$_2$ between the measured $\kappa_0/T$ and the value estimated via 
Eq.~4 would be entirely fortuitous in a nodal $s$-wave state.


\begin{figure}[t]
\centering
\includegraphics[width=8.5cm]{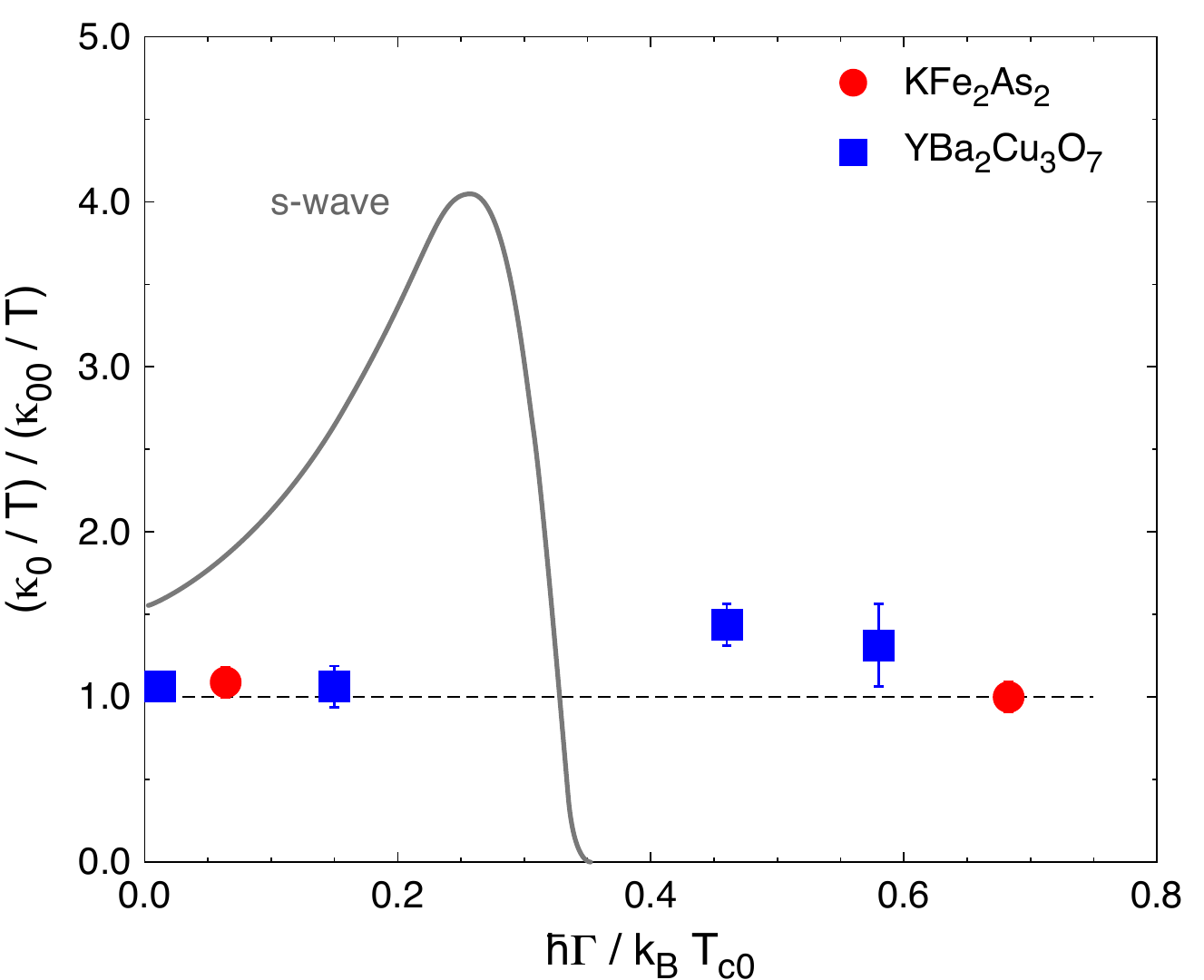}
\caption{
Dependence of $\kappa_0/T$ and on impurity  scattering rate $\Gamma$,
normalized by $T_{c0}$, the disorder-free superconducting temperature. 
Data for the pnictide KFe$_2$As$_2$ (red circles; from~\cite{Reid2012}) and the cuprate YBa$_2$Cu$_3$O$_7$ (blue squares; from~\cite{Taillefer1997}) are
normalized by the theoretically expected value for a $d$-wave superconductor, $\kappa_{00}/T = 3.3$ and 0.16 mW/K$^2$~cm, respectively (see text).
The typical dependence expected of an $s$-wave state with accidental nodes is also shown, 
from a calculation applied to pnictides (black line; from~\cite{Mishra2009}).
}
\label{Fig6}
\end{figure}


\subsection{Dependence of $\kappa_0/T$ on impurity scattering}  

A powerful test of whether line nodes are imposed by symmetry, as in a $d$-wave superconductor, or accidental,
as in an extended $s$-wave state, is the dependence of $\kappa_0/T$ on impurity (or defect) scattering.
In Fig.~6, we see that $\kappa_0/T$ in KFe$_2$As$_2$ is unchanged by a 10-fold increase in $\Gamma$ ({\it i.e.} a 
10-fold increase in $\rho_0$)~\cite{Reid2012}.
The classic data on the $d$-wave superconductor YBa$_2$Cu$_3$O$_7$~\cite{Taillefer1997},
reproduced here for comparison, show a similar insensitivity to disorder.
More than any other measurement, this rules out the possibility of an accidental line node in KFe$_2$As$_2$.
Calculations based on a two-band model for the pnictides produce a wide variety of curves for
$\kappa_0/T$ vs $\Gamma$~\cite{Mishra2009}; a typical one is reproduced in Fig.~6.
The value in the clean limit is different from the $d$-wave $\kappa_{00}/T$, by an amount which depends on parameters.
In general, $\kappa_0/T$ will depend strongly on $\Gamma$.
Here $\kappa_0/T$ starts by growing, as $v_\Delta$ falls with increasing $\Gamma$, 
then it goes through a near-divergence as $v_\Delta \to 0$ just before the nodes lift and finally it plummets to zero when
 the nodes are eventually lifted by sufficiently strong scattering
(see Fig.~6).

It is probably not possible, and certainly artificial, to adjust the various parameters of a two-band (or N-band) model 
of $s$-wave superconductivity in the pnictides to ensure that the following four facts are simultaneously reproduced:
1) correct value of $\kappa_0/T$;
2) constancy of $\kappa_0/T$ vs scattering up to $\Gamma/\Gamma_c = 0.5$;
3) suppression of $T_c$ to zero at $\Gamma_c$;
and 4) correct value of $\Gamma_c$ ({\it i.e.} $\simeq T_{c0}$).
As we saw, these four properties are naturally obtained in a $d$-wave state,
with no adjustable parameter.

\subsection{Dependence of $\kappa/T$ on temperature}  

Another test of $d$-wave theory applied to KFe$_2$As$_2$ is the temperature dependence of $\kappa/T$ at low temperature
($T \ll T_c$).
As seen in Fig.~5, the data for KFe$_2$As$_2$ obey
\begin{equation}
\frac{\kappa}{T} ~ = ~ \frac{\kappa_{0}}{T} ~~( ~1 + A~T^2 ~)~~~~~~~~~,
\end{equation}
with $A = 23$~K$^{-2}$.

Before we attribute this $A T^2$ term to electronic excitations, we need to know what is the phonon contribution.
Indeed, while phonons do not contribute to the residual linear term, they do of course carry heat, and
$\kappa_p \propto T^\alpha$, with $\alpha$ typically between 2 and 3~\cite{Sutherland2003,Li2008}.
There are several ways to show that the phonon term $\kappa_p$ is at least 10 times smaller than 
the $A T^2$ term below 0.4~K, and so cannot account for it.

The maximal value of the phonon conductivity $\kappa_p$ at sub-Kelvin temperatures is achieved 
when the phonon mean free path reaches the size of the sample. 
If that boundary scattering is diffuse (on rough surfaces), the $T$ dependence will be $\kappa_p = B T^3$, or  $\kappa_p/T = B T^2$~\cite{Li2008}. 
Sutherland {\it et al}. have calculated the expected phonon term $\kappa_p$ from the known phonon properties of the iron-based superconductor
LaFePO (specific heat and velocity) and the given sample dimensions~\cite{Sutherland2012}. 
Since their sample dimensions are comparable to ours and the sound velocities of iron-based materials are comparable, 
their estimate also applies approximately to our KFe$_2$As$_2$ sample. 
They obtain $B = 1.0$~mW/K$^4$~cm, while our measured slope is 100~mW/K$^4$~cm. 
There is no way that such a huge slope could be due to phonons. 
It must therefore be due to electronic excitations.

A second way to rule out phonons as the cause of the $T^2$ slope is to compare with a sample that has 10 times more impurity scattering. 
This should not affect $\kappa_p$ at those temperatures, where the phonon mean free path is controlled by the sample boundaries~\cite{Li2008}.
In Fig.~7, we see that it does.
Indeed, the slope of $\kappa/T$ drops by at least a factor 10.
A third way is to look at the data on Ba$_{0.6}$K$_{0.4}$Fe$_2$As$_2$, in Fig.~5.
Because Ba$_{0.6}$K$_{0.4}$Fe$_2$As$_2$ has no nodes, electrons do not contribute to the slope of $\kappa/T$ either (for $T < T_c / 100$).
We see that $\kappa_p \simeq B T^2$, and $B$ is 10 times smaller than the slope $A$ in KFe$_2$As$_2$.

Having shown that the $A T^2$ term in KFe$_2$As$_2$ is electronic, and hence due to nodal quasiparticles,
we have yet another, independent confirmation of $d$-wave behavior.
As we saw, according to Eq.~5, a $T^2$ dependence is the expected low-temperature behavior.
A $T^2$ dependence was indeed observed in YBa$_2$Cu$_3$O$_7$, with $\kappa/T = (\kappa_0/T) (1+19~T^2)$~\cite{Hill2004}.
This is yet another difference with respect to an extended $s$-wave state, for which calculations do not obtain a 
quadratic growth in $\kappa/T$ at low temperature~\cite{Mishra2009}.



\begin{figure}[t]
\centering
\includegraphics[width=8.5cm]{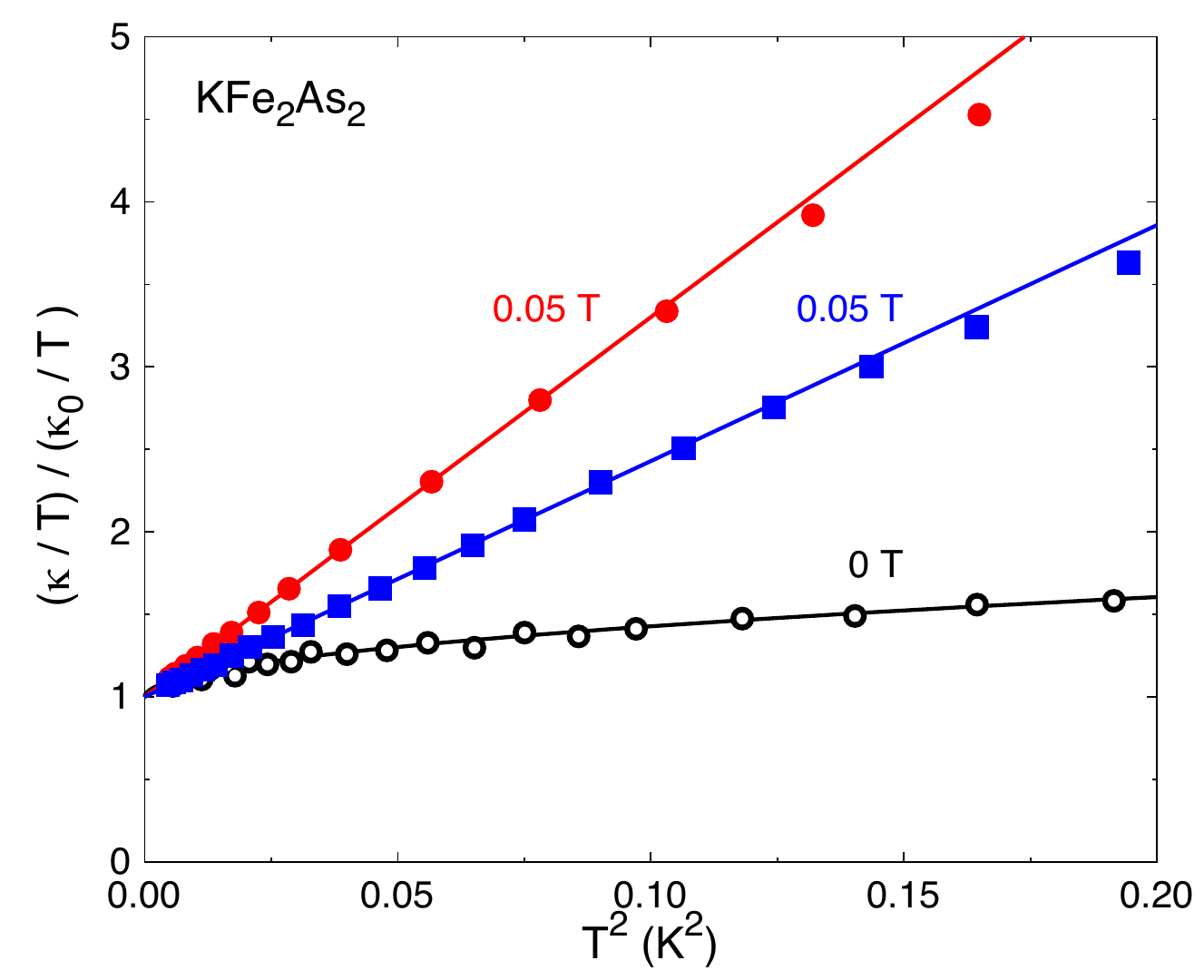}
\caption{
Thermal conductivity of KFe$_2$As$_2$ for a clean sample (closed symbols), whose $\rho_0$ increased by a factor 1.15 over
a period of 6 months (before, red circles~\cite{Reid2012}; after, blue squares), and a dirty sample (open black circles~\cite{Dong2010}), 
whose $\rho_0$ is 10 times higher than that of the clean sample.
The data is plotted as $\kappa/T$ vs $T^2$, normalized by the value ($\kappa_0/T$) in the $T=0$ limit.
Note that the actual value of $\kappa_0/T$ is identical in all three data sets, within error bars (at $H \to 0$).  
These data show that the $T^2$ slope is strongly reduced by an increase in impurity or defect scattering, as expected of a $d$-wave superconductor
(see Eq.~5).
}
\label{Fig6}
\end{figure}


Moreover, in $d$-wave theory the coefficient $A$ is expected to decrease with scattering, as we just saw it does.
In the limit of unitary scattering, $A \propto 1 / \Gamma$ , so that a 10-times larger $\Gamma$
would yield a 10-times smaller slope~\cite{Graf1996}, as roughly observed.
We can be more precise by using data on a high-quality sample whose RRR dropped slightly over a 6-month period,
by a factor of 1.15.
The data is shown in Fig.~7. 
The increased disorder scattering did not change $\kappa_0/T$ within error bars, but it did change the $T^2$ slope noticeably,
by a factor 1.60.
So we see again, now within the same sample, that an increase in $\Gamma$ causes a drop in the $T^2$ slope while
keeping $\kappa_0/T$ constant, as expected of a $d$-wave superconductor.

The temperature below which the $T^2$ dependence of $\kappa/T$ in KFe$_2$As$_2$ sets in, $T \simeq 0.1~T_c$,
is a measure of the impurity bandwidth $\gamma$.
It is in excellent agreement with the temperature below 
which the penetration depth $\lambda_a(T)$ of KFe$_2$As$_2$ (in a sample with similar RRR) 
deviates from its linear $T$ dependence~\cite{Hashimoto2010a}.
This is beautifully consistent with the theory of $d$-wave superconductors~\cite{Hirschfeld1993}.


\begin{figure}[t]
\centering
\includegraphics[width=8.5cm]{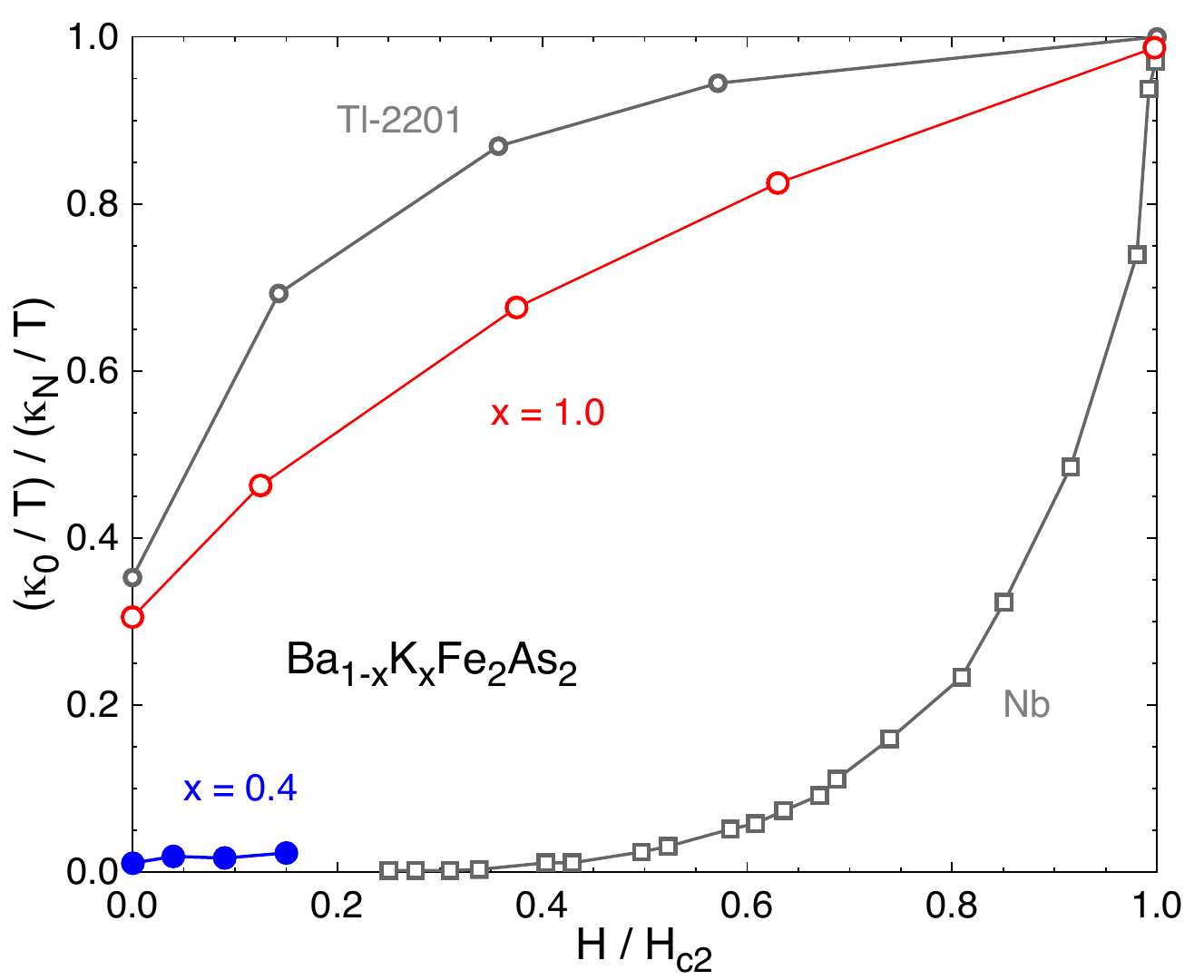}
\caption{
Residual linear term, $\kappa_0/T$, normalized by the normal-state thermal conductivity in the $T=0$ limit,
$\kappa_{\rm N}/T$, as a function of magnetic field $H$ applied along the $c$ axis,
normalized by the upper critical field $H_{c2}$, for two concentrations
of  Ba$_{1-x}$K$_x$Fe$_2$As$_2$:
$x=0.4$ (closed blue circles) \cite{Reid2011} and $x=1.0$ (open red circles) \cite{Dong2010}.
The values of $H_{c2}$ used here are $100$~T and $0.8$~T, respectively.
Data on KFe$_2$As$_2$ are from a sample in the dirty limit~\cite{Dong2010}.
For comparison, we reproduce corresponding data for the isotropic $s$-wave superconductor Nb
(open grey squares~\cite{Shakeripour2009})
and the dirty $d$-wave superconductor Tl-2201 (open grey circles~\cite{Proust2002}). 
}
\label{Fig3}
\end{figure}


\subsection{Dependence of $\kappa_0/T$ on magnetic field}  

The presence of nodes in the gap will cause $\kappa_0/T$ to be immediately enhanced by application of a magnetic field $H$.
A rapid initial rise in $\kappa_0/T$ vs $H$ has been observed in the cuprate $d$-wave superconductors
 YBa$_2$Cu$_3$O$_7$~\cite{Chiao1999} and Tl$_2$Ba$_2$CuO$_{6+\delta}$~\cite{Proust2002}, for example.
In the dirty limit, measurements on Tl$_2$Ba$_2$CuO$_{6+\delta}$ (with  $T_c = 13$~K)~\cite{Proust2002},
reveal a smooth monotonic increase of $\kappa_0/T$ vs $H$ all the way to $H_{c2} \simeq 7$~T, as reproduced in Fig.~8.
In dirty-limit samples, KFe$_2$As$_2$~\cite{Dong2010} and Tl$_2$Ba$_2$CuO$_{6+\delta}$~\cite{Proust2002} show 
very similar curves of $\kappa_0/T$ vs $H/H_{c2}$ (Fig.~8).
Note that the similar value of $(\kappa_0/T)/(\kappa_{\rm N}/T) \simeq 1/3$ at $H=0$ reveals a similar level of scattering, with $\hbar \Gamma / \Delta_0 \simeq 1/3$.

In the clean limit, one expects $\kappa_0/T$ to fall abruptly as the field is reduced below $H_{c2}$.
This is true for any type of gap. It comes from the fact that vortices enter the sample and immediately 
introduce an extra scattering process, limiting the long quasiparticle mean free path of the normal state.
In Fig.~8, this behavior is seen in the classic $s$-wave superconductor Nb.
A behavior almost identical to that of Nb was observed in clean-limit samples of the iron pnictide LiFeAs~\cite{Tanatar2011}, 
showing this particular stoichiometric pnictide to have an $s$-wave gap.
(At $H=0$, it also has $\kappa_0/T = 0$.)

Fig.~9 shows how theoretical calculations for a clean $d$-wave superconductor contain these two features:
immediate initial rise at low $H$ and sharp drop just below $H_{c2}$~\cite{Vekhter1999}.
Data on a clean sample of KFe$_2$As$_2$ behave precisely in this way, and are in remarkably good agreement with
$d$-wave calculations (see Fig.~9).
In the case of cuprates, a full comparison with $d$-wave theory has not so far been possible in the clean limit.
Although the clean limit is certainly reached in 
YBa$_2$Cu$_3$O$_7$, measurements have been limited to $H \ll H_{c2} $.

In Fig.~8, the data for  Ba$_{0.6}$K$_{0.4}$Fe$_2$As$_2$ are also shown, and the increase in $\kappa_0/T$ at low field is seen to be very flat~\cite{Reid2011}.
This is further evidence that there are no nodes in the gap, and it also shows that there are no deep minima.
Because the same flat dependence is seen for a heat current along the $c$ axis, it reveals a full isotropic gap at $x=0.4$~\cite{Reid2011}.


\begin{figure}[t]
\centering
\includegraphics[width=8.5cm]{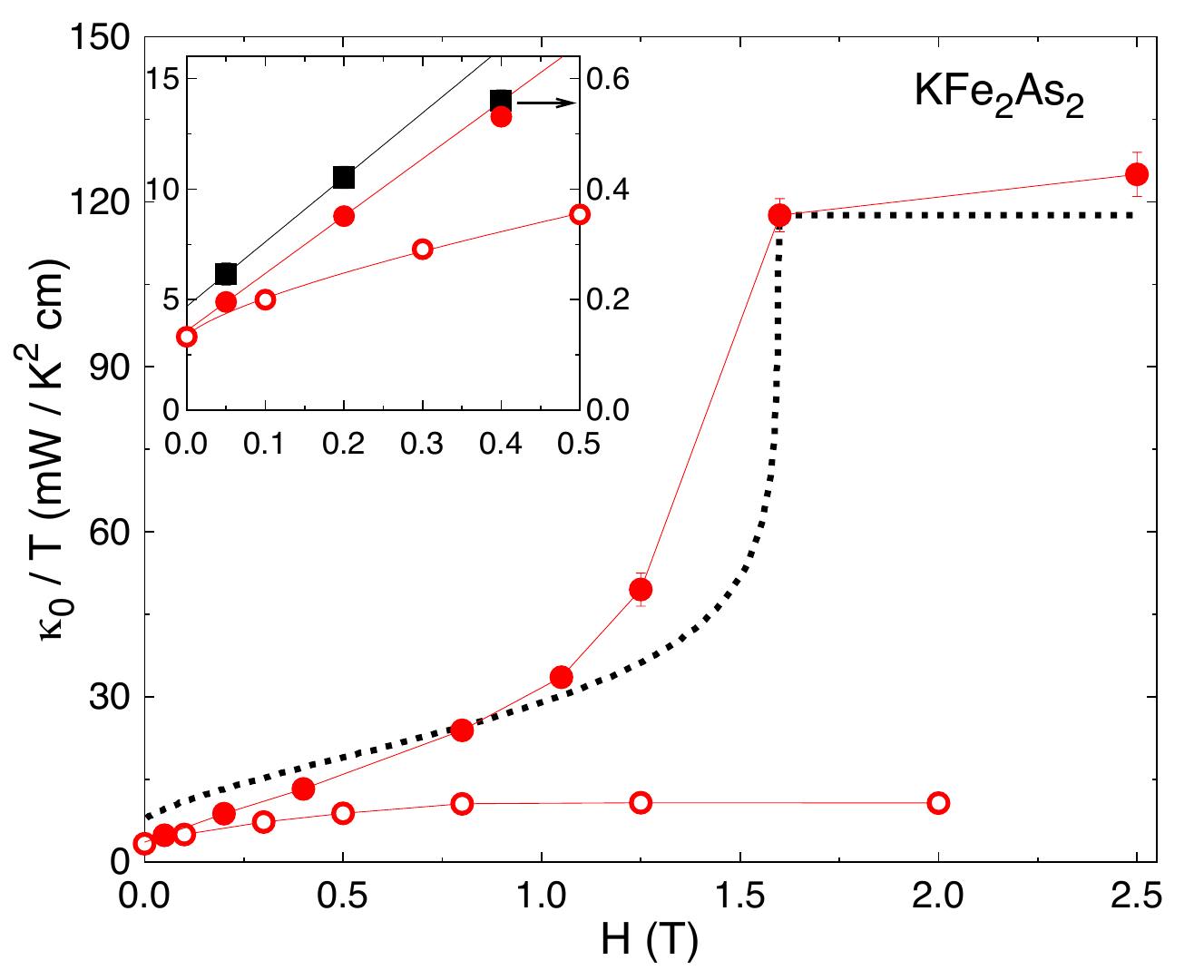}
\caption{
Field dependence of $\kappa_0/T$ measured in two samples of KFe$_2$As$_2$ for a current parallel to the $a$ axis ($J || a$)
and a field parallel to the $c$ axis ($H || c$):
one in the clean limit ($\Gamma/\Gamma_c = 0.05$; full red circles \cite{Reid2012}),
the other in the dirty limit ($\Gamma/\Gamma_c = 0.5$; open red circles \cite{Dong2010}).
The black dashed line is a theoretical calculation for a $d$-wave superconductor in the clean limit ($\hbar \Gamma/\Delta_0 = 0.1$)~\cite{Vekhter1999}. 
{\it Inset}: 
Zoom of the same data at low field (left axis).
Also shown is $\kappa_{0}/T$ measured for a current along the $c$ axis ($J\parallel c$) in a sample with $\Gamma/\Gamma_c = 0.10$ 
(black squares~\cite{Reid2012}, right axis). 
%
}
\label{Fig4}
\end{figure}


\subsection{Dependence of $\kappa_0/T$ on current direction}  

Our assignment of a $d$-wave gap in KFe$_2$As$_2$ must withstand yet another stringent test: the anisotropy of the
gap structure. 
As sketched in Fig.~3, a $d$-wave gap on a single quasi-2D cylindrical Fermi surface (at the zone center) would 
necessarily have 4 line nodes that run "vertically" along the length of the cylinder (along the $c$ axis).  
In such a nodal structure, zero-energy nodal quasiparticles will conduct heat not only in the plane, 
but also along the $c$ axis,
by an amount proportional to the $c$-axis dispersion of the Fermi surface.
In the simplest case, $c$-axis conduction will be smaller than $a$-axis conduction by a factor equal to
the mass tensor anisotropy ($v_{\rm F}^2$ in Eq.~3).
In other words, $(\kappa_{c0}/T) / (\kappa_{a0}/T) \simeq (\kappa_{c{\rm N}}/T) / (\kappa_{a{\rm N}}/T) = (\sigma_{c} / \sigma_{a})_{\rm N}$,
the anisotropy in the normal-state thermal and electrical conductivities.

This is confirmed by calculations for a quasi-2D $d$-wave superconductor~\cite{Vekhter2007},
whose vertical line nodes yield an anisotropy ratio in the superconducting state very similar to that of the normal state.
%


\begin{figure}[t]
\centering
\includegraphics[width=8.5cm]{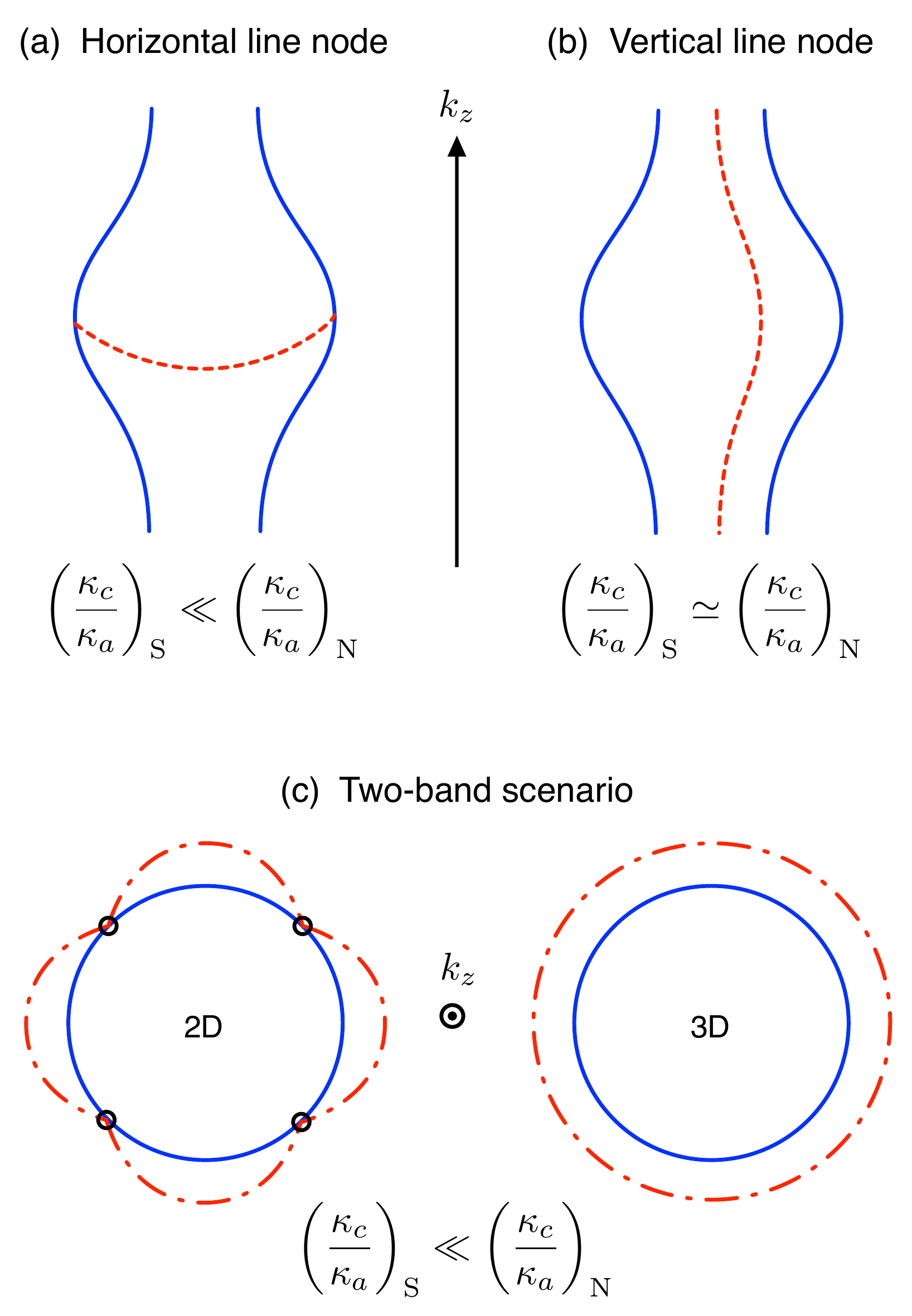}
\caption{
Sketch of different scenarios for line nodes on the Fermi surface of \KFeAs\ ($x=1$).
In panels (a) and (b), the effect of a line node on the anisotropy of heat transport is illustrated in a single-band model,
whereby  the Fermi surface is approximated by a single warped cylinder.
The $c$-axis dispersion (along $k_z$) is responsible for conduction along the $c$ axis, producing 
an $a$-$c$ anisotropy in the normal-state (N) conductivity $\sigma$, 
$(\sigma_c / \sigma_a)_{\rm N}$, which in KFe$_2$As$_2$ is equal to 0.04~\cite{Reid2012}.
In the normal state at $T \to 0$, the very same anisotropy is observed in the thermal conductivity $\kappa$,
so that $(\kappa_c / \kappa_a)_{\rm N} = 0.04$.
The onset of superconductivity will alter this anisotropy, depending on the anisotropy of the gap.
For a gap with a (horizontal) line node in the basal plane (a), 
thermal conduction in the basal plane will be much better than conduction along the $c$ axis, so that in the
superconducting state (S) in zero field, 
 $(\kappa_c / \kappa_a)_{\rm S} \ll (\kappa_c / \kappa_a)_{\rm N}$.
 By contrast, if the line node is vertical, lying within the $a$-$c$ plane, then the anisotropy
 of the normal state is more or less preserved, and
$(\kappa_c / \kappa_a)_{\rm S} \simeq (\kappa_c / \kappa_a)_{\rm N}$~\cite{Vekhter2007}.
In (c), we illustrate how in a two-band scenario, a vertical line node present only on one of the two Fermi surfaces
can also lead to a much reduced anisotropy in the superconducting state. This will happen if $c$-axis conduction 
in the normal state is caused predominantly by that surface (with strong 3D character) which does not have the accidental line node.   
}
\label{Fig5}
\end{figure}


This is what we see in KFe$_2$As$_2$ (inset of Fig.~9):
($\kappa_{a0}/T) / (\kappa_{c0}/T) = 20 \pm 4$,
 very close to the intrinsic normal-state anisotropy $(\sigma_{a} / \sigma_{c})_{\rm N} = 25 \pm 1$~\cite{Reid2012}.
This shows that the nodes in KFe$_2$As$_2$ are vertical lines running along the $c$ axis.

If there were only a single quasi-2D sheet in the Fermi surface of KFe$_2$As$_2$, the anisotropy could equally be due to 
symmetry-imposed vertical lines in a $d$-wave gap or accidental vertical line nodes in an extended $s$-wave gap.
However, the fact that the Fermi surface of KFe$_2$As$_2$ contains several sheets with very different $c$-axis dispersions
provides compelling evidence in favor of $d$-symmetry. 
The argument is as follows.
As sketched in Fig.~10c, if one Fermi surface sheet dominates the $c$-axis transport in the normal state
because it has the strongest $c$-axis dispersion,
but it does not have nodes in the superconducting state, then the anisotropy $(\kappa_{a} / \kappa_{c})_{\rm S}$ measured in the superconducting state
will be much larger than the anisotropy $(\kappa_{a} / \kappa_{c})_{\rm N}$ in the normal state.
Now the Fermi surface of KFe$_2$As$_2$ does contain a small pocket at the $Z$ point whose strong 3D character~\cite{Hashimoto2010a,Yoshida2012} 
will make it a dominant contributor to $c$-axis conduction. 
In an extended $s$-wave scenario, there would be no particular reason why the gap would develop nodes on this small pocket,
and so the anisotropy would be very different in the superconducting and normal states, unlike what is measured.
By contrast, in $d$-wave symmetry this pocket would automatically and necessarily have nodes, as any other zone-centered sheet,
thereby ensuring that transport anisotropy remains similar in the superconducting and normal states.
The same argument applies to the three $\Gamma$-centered cylinders, one of which has a much stronger $c$-axis dispersion than the other two~\cite{Hashimoto2010a,Yoshida2012}.
Again, all three must have vertical line nodes to preserve the similarity of anisotropies in superconducting and normal states.
This is automatically ensured by $d$-wave symmetry,
but is not the case in $s$-wave scenarios such as that of ref.~\cite{Maiti2012}, where only one of the three sheets has nodes.

\section{LaFePO: a nodal $s$-wave superconductor}  

In light of what we know of KFe$_2$As$_2$, it is interesting to investigate the case  
of the iron-based superconductor LaFePO.
In some ways, LaFePO shares several similarities with KFe$_2$As$_2$: it is stoichiometric, with a relatively small 
$T_c \simeq 7.5$~K, and its gap has nodes, as detected by a linear $T$ dependence of the penetration depth~\cite{LaFePO-lambda}
and a residual linear term in the zero-field thermal conductivity~\cite{LaFePO-kappa}.
Are those line nodes imposed by symmetry or accidental? 
Is the symmetry $d$-wave or $s$-wave?

These questions were addressed in a recent study of heat transport in LaFePO~\cite{Sutherland2012},
where some of the aspects considered here were covered.
%
Two features are singled out by the authors as inconsistent with $d$-wave symmetry. 
First, $T_c$ seems to be robust against impurity scattering.
Although only one sample was studied, the estimated scattering rate for that sample
is such that $\hbar \Gamma \simeq k_{\rm B} T_{c}$.
$d$-wave superconductivity would be almost entirely suppressed by such a large scattering rate.
It doesn't appear to be. This robustness certainly contrasts with the high sensitivity of KFe$_2$As$_2$,
and is more reminiscent of optimally-doped BaFe$_2$As$_2$, SrFe$_2$As$_2$ and CaFe$_2$As$_2$~\cite{Kirshenbaum2012}.
Note that a systematic study of the impact of impurity scattering on the $T_c$ of LaFePO is needed to confirm this claim.

The second feature is the temperature dependence of the electronic part of $\kappa/T$, which is linear at 
low temperature ($T < T_c / 15$)~\cite{Sutherland2012} rather than $T^2$ as in KFe$_2$As$_2$ and YBa$_2$Cu$_3$O$_7$.
These two facts suggest that an extended $s$-wave scenario is more likely for LaFePO,
in agreement with renormalization-group calculations~\cite{Wang2010,Thomale2011b}.
A good test would be to study the dependence of $\kappa_0/T$ on $\Gamma$ and on current direction.

Note that the iron pnictides BaFe$_2$(As$_{1-x}$P$_x$)$_2$~\cite{Hashimoto2010} and Ba(Fe$_{1-x}$Ru$_x$)$_2$As$_2$~\cite{Qiu2011} 
also have nodes in the gap. 
The robustness of $T_c$ to impurity scattering in the latter~\cite{Kirshenbaum2012} is again an argument in favor of a nodal $s$-wave gap.

\section{Summary}  

In this section, we summarize the evidence presented in this Article that leads us to identify the pairing symmetry 
of the stoichiometric iron pnictide superconductor KFe$_2$As$_2$ as $d$-wave, and why the 
pairing symmetry of Ba-doped KFe$_2$As$_2$ must be different. 
It is impressive that the respective $d$-wave and $s$-wave symmetries of these two otherwise quite similar materials
was correctly predicted by renormalization-group calculations~\cite{Thomale2011}.

\subsection{Ba$_{0.6}$K$_{0.4}$Fe$_2$As$_2$: an $s$-wave superconductor}  

The fact that $\kappa_0/T = 0$ at $H=0$ in Ba$_{0.6}$K$_{0.4}$Fe$_2$As$_2$ for current directions parallel and perpendicular
to the $c$ axis~\cite{Reid2011} shows that there are no nodes in the gap of this material,
immediately ruling out $d$-wave pairing.
The very slow increase of $\kappa_0/T$ with field, again in both current directions~\cite{Reid2011}, also reveals that the gap has no deep minima
and is essentially isotropic. 
This suggests $s$-wave symmetry, with or  without a change of sign between the hole and electron Fermi surfaces.
The striking fact about this superconductor remains its high critical temperature, $T_c \simeq 40$~K, and its very
high upper critical field, $H_{c2} \simeq 100$~T~\cite{Hc2A,Hc2B}.

\subsection{KFe$_2$As$_2$: a $d$-wave superconductor}  

The six independent experimental properties that confirm KFe$_2$As$_2$ as a $d$-wave superconductor
are listed in Table~I. We summarize them here.
$T_c$ is suppressed by impurity scattering at a rate consistent with $\hbar \Gamma_c / k_{\rm B} T_{c0} \simeq 1.0$.
This rapid rate, precisely as expected for $d$-wave, is 30 to 50 times more rapid than what is observed
in other iron-pnictide superconductors with the same crystal structure. This is a huge difference,
pointing compellingly to a difference in pairing symmetry.
Nodal quasiparticles have a conductivity which is unaffected by a 10-fold increase in impurity scattering.
This universal heat conduction is the signature of line nodes imposed by symmetry.
The magnitude of that conductivity and its $a$-$c$ anisotropy are precisely those expected of a $d$-wave gap,
with vertical line nodes on all the zone-centered Fermi surfaces.
The initial growth in $\kappa/T$ with temperature is precisely $T^2$, 
precisely as expected of a $d$-wave superconductor.
Moreover, the $T^2$ term is rapidly suppressed by scattering, again as expected.
The entire field dependence of $\kappa_0/T$, from $H=0$ to $H = H_{c2}$, is in striking agreement with calculations
for a $d$-wave gap.

By contrast, an $s$-wave gap with accidental line nodes on one Fermi surface cannot account for all 
six of these features simultaneously. 
%
Impurity scattering will in general have a strong effect on $\kappa_0/T$ and it is known to have a
weak effect on $T_c$ in $s$-wave superconductors of the 122 structure --
the opposite of what is seen in KFe$_2$As$_2$.
The magnitude of $\kappa_0/T$ in clean samples will only be right by accident, and only by another coincidence 
will the $a$-$c$ anisotropy be right.
The $T$ dependence will not in general be the observed $T^2$.

We may now ask how the change in order parameter symmetry from $s$-wave at $x=0.4$ to $d$-wave at $x=1.0$ proceeds as a function
of $x$ in Ba$_{1-x}$K$_{x}$Fe$_2$As$_2$.
One scenario is two separate superconducting phases, separated by a non-superconducting region.
%
A second scenario would be to have a continuous superconducting phase where the change in symmetry from $s$-wave to $d$-wave proceeds via an 
intermediate state of $\p{s \pm id}$ symmetry, a complex order parameter that breaks time-reversal symmetry~\cite{Platt2012}. 
%
Further work is required to elucidate this question.


\begin{table}[t]
\caption{
Quantitative and qualitative characteristics of superconductivity in KFe$_2$As$_2$, compared to the corresponding
characteristics expected of a $d$-wave superconductor (with symmetry-imposed vertical line nodes)
and of an extended $s$-wave superconductor (with accidental vertical line nodes).
$\Gamma_c$ is the critical impurity scattering rate needed to suppress $T_c$ to zero.
$T_{c0}$ is the clean-limit value of the superconducting transition temperature.
$\kappa_0/T$ is the residual linear term in the thermal conductivity extrapolated to $T=0$ and $H=0$.
$\kappa_{00}/T \equiv  \hbar\gamma_{\rm N}v_{\rm N}^2 / (4.28\pi k_{\rm B} T_{c0})$.
$\p{\kappa_c/\kappa_a}_{\rm S}$ is the anisotropy of heat conduction in the superconducting state
at $T \to 0$, namely the ratio of $\kappa_0/T$ measured along the $c$ axis over
$\kappa_0/T$ measured along the $a$ axis.
$\p{\kappa_c/\kappa_a}_{\rm N}$ is the corresponding anisotropy of heat conduction in the normal state,
equal to $\p{\sigma_c/\sigma_a}_{\rm N} $, the anisotropy in the electrical conductivity $\sigma$.
\\
$^\star$ Extended s-wave with accidental vertical line nodes.\\
$\dagger$~Value obtained in optimally-doped $AE$Fe$_2$As$_2$ (see Fig.~4 and~\cite{Kirshenbaum2012}). \\
$\ddagger$ Strongly dependent on details - can be much more or much less than 1.0.

}

\begin{center}
\begin{tabular}{l c c c}
												&~~~KFe$_2$As$_2$~~~		&~~~$d$-wave~~~		&~~~$s$-wave$^\star$~~~\\
\toprule

$\hbar \Gamma_c / k_{\rm B} T_{c0}$							&$1.3\pm 0.2$				&0.88			&~~45~$\dagger$	\\
$\p{\kappa_0/T} / \p{\kappa_{00}/T}$							&$1.1 \pm 0.5$				&1.0				& $\ddagger$		\\
$\p{\kappa_c/\kappa_a}_{\rm S} / \p{\kappa_c/\kappa_a}_{\rm N} $	&1.25					&$\simeq$~1.0~~~	& $\ddagger$			\\ 
\midrule
$\Gamma$ dependence of $\kappa_0/T$                   				& weak					& weak			& strong				\\
$T$ dependence of $\kappa/T$								& $T^2$					& $T^2$			& $T$					\\
$H$ dependence of $\kappa_0/T$ 							& rapid					& rapid			&rapid		\\
\bottomrule
\end{tabular}
\label{table1}
\end{center}
\end{table}



\section*{Acknowledgements}  

We thank A.~Carrington, A.~Chubukov, R. Fernandes, R.~W.~Hill, P.~J.~Hirschfeld, J. Paglione, S.~Y.~Li, M.~Sutherland, R.~Thomale and I.~Vekhter for fruitful discussions and J. Corbin for his assistance with the experiments. 
The work at Sherbrooke was supported by a Canada Research Chair, CIFAR, NSERC, CFI and FQRNT.
The work at Ames was supported by the U.S. Department of Energy, Office of Basic Energy Sciences, 
Division of Materials Sciences and Engineering under contract No. DE-AC02-07CH11358.
The work in Japan was supported by Grants-in-Aid for Scientific Research 
(Nos. 21540351 \& 22684016) from MEXT and JSPS and 
Innovative Areas ''Heavy Electrons" (Nos. 20102005 \& 21102505) from MEXT,
Global COE and AGGST financial support program from Chiba University. 
The work in China was supported by NSFC and the MOST of China (\#2011CBA00100).

\textit{Note added in proof} - By adding Co impurities in KFe$_2$As$_2$, a recent study~\cite{Wang2012} has confirmed that $T_c$ falls rapidly to zero with impurity scattering, roughly at $\rho_{crit} = 4.5~\mu \Omega$~cm, and the residual linear term in the thermal conductivity is indeed universal, remaining approximately constant even when the normal-state conductivity is decreased by a factor 30.

\end{document}